\documentclass[aps,prl, twocolumn,groupedaddress]{revtex4-1}
\usepackage{graphicx}
\usepackage{amssymb}
\usepackage{amsmath}
\usepackage{amsthm}
\usepackage{hyperref}
\usepackage{epstopdf}
\usepackage{tikz}
\usepackage{comment}
\usepackage{appendix}
\usepackage{lipsum}
\usepackage{psfrag}
\usepackage{pstool}
\usepackage{caption}

\DeclareMathOperator{\e}{e}
\def\dd{{\rm d}}
\def\ii{{\rm i}}

\def\be{\begin{equation}}
\def\ee{\end{equation}}
\def\bea{\begin{eqnarray}}
\def\eea{\end{eqnarray}}

\def\a{\alpha}
\def\b{\beta}

\begin{document}


\title{Exact confirmation of 1D nonlinear fluctuating hydrodynamics \\ for a two-species exclusion process}


\author{Zeying Chen}
\email[]{zeyingc@student.unimelb.edu.au}

\author{Jan de Gier}
\email[]{jdgier@unimelb.edu.au}
\affiliation{ARC Centre of Excellence for Mathematical and Statistical Frontiers (ACEMS), School of Mathematics and Statistics, The University of Melbourne, VIC 3010, Australia}

\author{Iori Hiki}
\email[]{hiki.i@stat.phys.titech.ac.jp}

\author{Tomohiro Sasamoto}
\email[]{sasamoto@phys.titech.ac.jp}

\affiliation{Department of Physics, Tokyo Institute of Technology, Ookayama 2-12-1, Tokyo 152-8551, Japan}

\thanks{The authors are grateful to M.~Wheeler for suggesting an idea leading to the formula (5) and to P.L.~Ferrari, I.~Kostov, H.~Spohn for discussions. This work was initiated at KITP Santa Barbara for during the program \textit{New approaches to non-equilibrium and random systems: KPZ integrability, universality, applications and experiments}, which was supported in part by the National Science Foundation under Grant No. NSF PHY11-25915. Part of this work was performed during a stay of all authors at the MATRIX mathematical research institute in Australia during the program \textit{Non-equilibrium systems and special functions}. JdG and ZC gratefully acknowledge support from the Australian Research Council. The work of TS is supported by JSPS KAKENHI Grants No. JP25103004, No. JP15K05203, No. JP16H06338.} 

\date{\today}

\begin{abstract}
We consider current statistics for a two species exclusion process of particles hopping in opposite directions on a one-dimensional lattice. We derive an exact formula for the Green's function as well as for a joint current distribution of the model, and study its long time behavior. For a step type initial condition, we show that the limiting distribution is a product of the Gaussian and the GUE Tracy-Widom distribution. This is the first analytic confirmation for a multi-component system of a prediction from the recently proposed non-linear fluctuating hydrodynamics for one dimensional systems.  
\end{abstract}


\maketitle

Macroscopic evolution of many physical particle systems is described by a hydrodynamic theory. To describe fluctuations and correlations it is customary to add noise to a linearized equation resulting in the theory of fluctuating hydrodynamics (FHD), see e.g.\cite{LanLif}. Despite its successes, FHD is often insufficient especially in low 
dimensions where anomalous transport is observed \cite{LLP2003,BBO2006,Dhar2008,Lepri2016} and one has to consider a nonlinear theory (NLFHD) \cite{MF1973, SH1977,DasMazenko1986}. 

NLFHD is in general difficult to handle, but in the case of one dimension important progress has been made through the connection to the Kardar-Parisi-Zhang (KPZ) equation \cite{KPZ1986,BS1995} (or the noisy Burgers equation) for which many exact results are known. Van Beijeren \cite{Beijeren2012} predicted that certain correlations of 
rather generic one-dimensional fluids in equilibrium may be captured by the KPZ equation in the stationary state 
\cite{PS2004, FS2006, IS2012,*IS2013}. Spohn reformulated this prediction on more general grounds \cite{Spohn2014}. 

The predictions in \cite{Beijeren2012,Spohn2014} were first mainly intended for one dimensional systems in equilibrium of
particles interacting via nonlinear potential and anharmonic chains such as the Fermi-Pasta-Ulam (FPU) chain 
\cite{FPU1955,Gallavotti2008},
which are governed by (deterministic) Hamiltonian dynamics. 
Similar predictions can also be formulated for stochastic dynamics \cite{FSS2013}, again originally for the stationary situation but they have since been extended to current distributions and for a transient regime from a step type initial condition \cite{MS2016,*MS2017}. 

These predictions of NLFHD have been tested in many numerical simulations 
\cite{DDSMS2014,MS2013,*MS2014,*MS2015,LBCCGL2018p,KHS2015,PSS2014,*PSS2015}
but our theoretical understanding of their validity is unsatisfactory because NLFHD is based on a heuristic decoupling of modes for which there has not been a firm analytic confirmation although some support is given by mode-coupling 
theory. It is highly desirable therefore to establish the correctness of NLFHD predictions for concrete microscopic models.

Such analytic confirmation would be difficult to achieve for Hamiltonian dynamics as it implies understanding of the long time behavior of a chaotic system. In the case of stochastic dynamics there is no such fundamental difficulty. In this Letter we provide the first confirmation from first principles of the predictions of NLFHD for a two-component stochastic system via exact results for the Green's function and a joint current distribution. 
 
For one dimensional systems with a single mode, notably those in the KPZ universality class, there has been remarkable progress in our understanding of their fluctuations over the last twenty years \cite{Corwin2012,QS2015}. 
Discoveries of several exact solutions for models in the universality class, such as the asymmetric simple exclusion process (ASEP) and the KPZ equation \cite{Johansson2000,TW2009a, SS2010a,*SS2010b,*SS2010c,ACQ2011}, have allowed us to study fluctuation properties in quite some detail. For example, it is possible to derive universal distribution functions and correlations of physical quantities in the scaling limit, which have turned out to show intriguing dependence on geometry and initial conditions
\cite{BR2000,PS2002a, Sasamoto2005,CLD2011}, as can also be observed in real experiments \cite{TS2010,*TS2012,TSSS2011,FT2017}.  

For systems with multiple modes progress has been much slower. For example, most studies based on exact solutions have 
so far been restricted to stationary properties \cite{BE2007,PEM2009,KMO2016,CEMRV2016} or the level of critical exponents \cite{ADHR1994, KimdenNijs2007, AKSS2009}. But there have been a few recent results for the Green's function \cite{TW2013,Kuan2018p} and there is a growing need for the extension of methods to derive distribution functions of physical quantities to multi-component systems \cite{Kuan2016, CdGW2017p,FNG2017p}. 

We report on the two species Arndt-Heinzl-Rittenberg (AHR) exclusion process \cite{AHR1999}, and give an exact multiple integral formula of a joint current distribution. For a certain mix of step and Bernoulli initial condition, we show that this distribution in the scaling limit tends to a product of Gaussian and GUE Tracy-Widom distribution
from random matrix theory \cite{TW1994, Mehta2004, Forrester2010}, as predicted by NLFHD. 

\begin{figure}
\scalebox{0.3}{\includegraphics{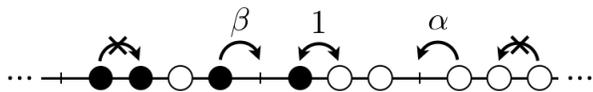}}
\caption{
The AHR model. 
A $+$ (resp. $-$) particle, denoted by $\bullet$ (resp. $\circ$), 
hops to the right (resp. left) with rate $\b$ (resp. $\a$). A pair of $+$ and $-$ on neighboring sites 
swap their positions with rate 1. }
\label{fig:AHR}
\end{figure}

The AHR model is a stochastic Markov process consisting of two families of particles that hop in opposite directions
on a one dimensional lattice. 
The transition rates for $+$ and $-$ particles in the AHR model are
\begin{eqnarray}
\b: &&\quad  (+,0) \rightarrow (0,+),\quad
\a:\quad (0,-) \rightarrow (-,0), \nonumber\\
1: &&\quad (+,-) \rightarrow (-,+) .
\end{eqnarray}
See FIG. \ref{fig:AHR}. The AHR model is known to be Yang-Baxter integrable \cite{Cantini2008}, and the stationary state of the AHR model on a ring can be given in terms of the matrix product form \cite{AHR1999, RSS2000}. Throughout we will take $\a+\b=1$ for which the stationary state is factorised. Moreover, later we will specialise to $\a=\b=1/2$.
We consider this AHR model on the infinite lattice.  

\paragraph{Green's function.} Our first main result is a formula for the Green's function of the model. We denote a configuration by the coordinates $x_1,\ldots,x_N$ of the $+$ particles and coordinates $y_1,\ldots,y_M$ of the $-$ particles. We impose an initial condition where all the $+$ particles are to the left of all the $-$ particles, which results in a simple structure.  Putting the superscript $(0)$ to represent the initial coordinates
we thus assume that $x_1^{(0)} < \ldots < x_{N}^{(0)} < y_1^{(0)} < \ldots < y_M^{(0)}$.

Let us denote by $G_{+-}$ the Green's function for the case where the ordering of the coordinates $x$ and $y$ is still the same as that of the initial condition, and by $G_{-+}$ the Green's function for the case where the two families have completely crossed after some large enough time $t$, i.e. when $y_1 < \ldots < y_M < x_1 < \ldots < x_{N}$. It can be shown 
that the Green's function for these cases is given by \cite{ChenGHS}
\begin{align}
&G_{\sigma\sigma'}(x,y,t|x^0,y^0) \nonumber\\
&= \oint \prod_{j=1}^N \frac{\dd z_j }{2\pi \ii} \prod_{k=1}^M \frac{\dd w_k }{2\pi \ii}\ \e^{\Lambda_{N,M} t}  S_{\sigma\sigma'} (\{z\},\{w\}) \nonumber\\
&\times \det\left( \left( \frac{z_j-1}{z_i-1}\right)^{j-1}  z_i^{x_j} \right)  \prod_{j=1}^N z_{j}^{-x^{(0)}_j-1} \nonumber \\
&\times \det\left( \left( \frac{w_k-1}{w_l-1}\right)^{m-k}  w_l^{-y_k} \right)  \prod_{k=1}^M w_{k}^{y^{(0)}_k-1},
\label{eq:G}
\end{align}
where all contours are around the origin, and where
\begin{align}
S_{+-} (\{z\},\{w\}) &=1, \nonumber\\
S_{-+} (\{z\},\{w\}) &= \prod_{k=1}^M \prod _{j=1}^{N} \frac{1}{\a z_{j} + \b w_{k}} ,
\end{align}
and $\Lambda_{N,M} = \b \sum_{i=1}^N (z_i^{-1}-1) +\a \sum_{i=1}^M (w_i^{-1}-1) $.
A formula for $G$ with a more complicated $S$ can also be derived for mixed positions of $+$ and $-$ particles. It is easy to check that \eqref{eq:G} satisfies the Markov dynamics of the AHR model and the initial condition. 
This type of formula was known for the single species TASEP \cite{Schuetz1997b}. 

The Green's function \eqref{eq:G} is found by constructing the eigenfunctions of the AHR Markov generator using a form of the Bethe ansatz analogous to that used for random tilings \cite{widom,gierN97,*gierN98}, which is related to the combinatorics of Littlewood-Richardson coefficients \cite{ZinnJustin2009,WZ2017}, i.e. we employ a representation of the eigenfunction in which two sets of variables $z_j$'s and $w_k$'s are associated with positive and negative particles respectively.  A more complicated formula would follow from the standard nested Bethe ansatz \cite{Cantini2008}.

\begin{figure}
\vspace{-3mm}
\scalebox{0.18}{\includegraphics{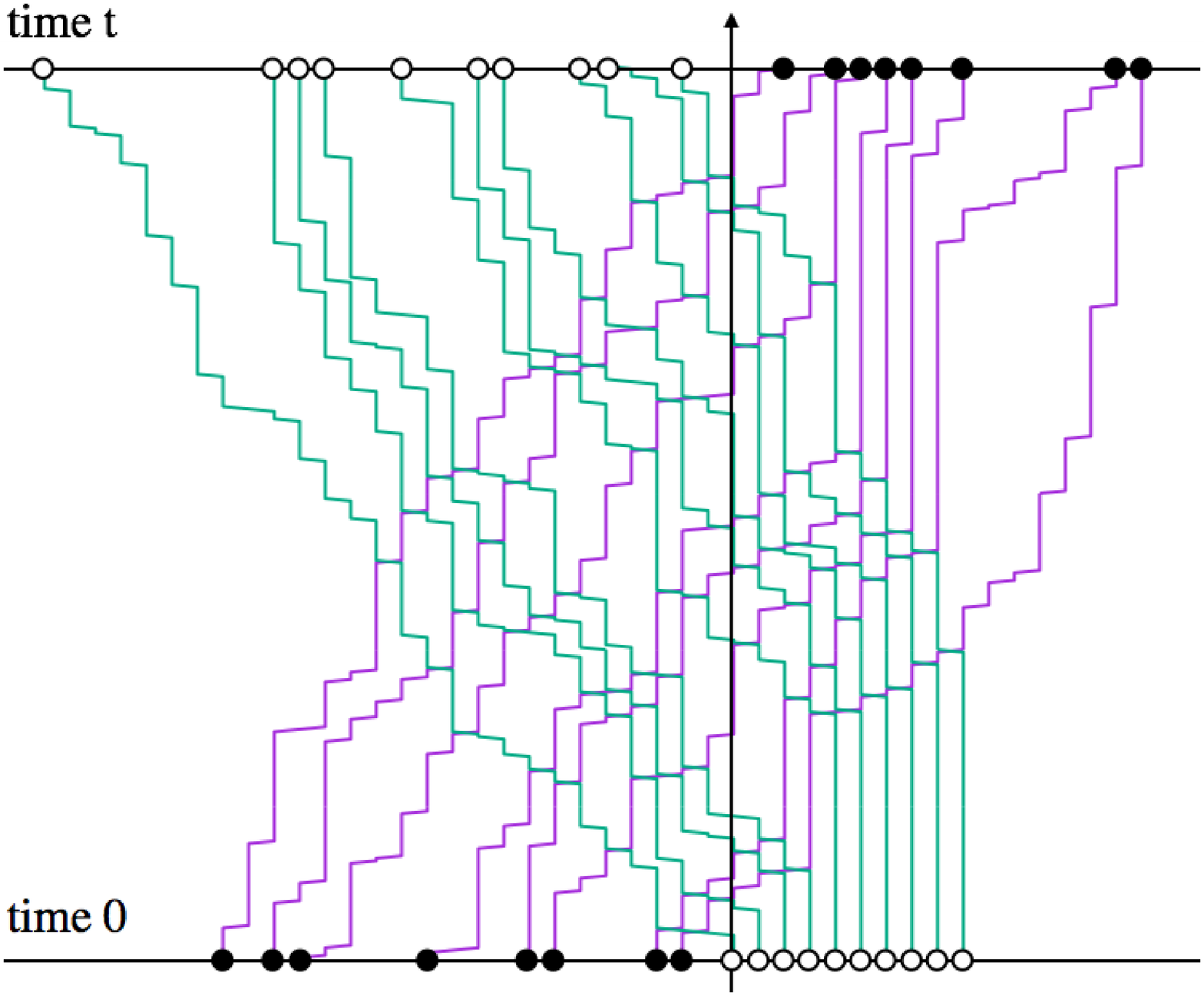}}
\caption{The initial condition consisting of $N$ $+$ particles ($\bullet$) with density $\rho$ on the left and 
$M$ $-$ particles ($\circ$) packed on the right. In this letter, we focus on the case of total exchange.}
\label{fig:AHRicex}
\end{figure}

\paragraph{Joint current distribution.} Our second main result is an exact expression for a joint current distribution for the AHR model. In the following we focus on the case in which initially $N$ particles of $+$ type are distributed by the Bernoulli measure with density $\rho$ on $x\leq -1$ and the first $M$ sites on $x\geq 0$ are occupied by $-$ particles. Let $n_\pm(t)$ denote the number of $\pm$ particles that have crossed the origin up to time $t$ and $P_{N,M,\rho}(t)=\mathbb{P}_{N,M}[n_+(t)=N,n_-(t)=M]$ the probability that all $+$ and $-$ particles have crossed the origin by time $t$. See FIG. \ref{fig:AHRicex}. 

The probability $P_{N,M,\rho}(t)$ can be written as a sum of the Green's function \eqref{eq:G} over all possible final positions of $+$ and $-$ particles, and also over the initial coordinates $x_j^{(0)}$ in which the distances among $+$ particles are independently distributed as a geometric random variable with parameter $1-\rho$.
After performing the associated geometric series we find that \cite{ChenGHS}
\begin{multline}
P_{N,M,\rho}(t)= \frac{1}{N!M!}
\oint \prod_{j=1}^N \frac{\dd z_j }{2\pi \ii} \prod_{k=1}^M \frac{\dd w_k }{2\pi \ii}\ \e^{\Lambda t} \times\\
\frac{\rho^N \prod_{1\le i<j\le N} (z_i-z_j)^2}{\prod_{j=1}^N (z_j-1)^{N} (1-(1-\rho) z_j) } \frac{ \prod_{1\le k<l\le M} (w_l-w_k)^2}{\prod_{k=1}^M (w_k-1)^{M} } \\
\times \frac{1} { \prod_{j=1}^N \prod_{k=1}^M \Big(\a z_j + \b w_k\Big)},
\label{eq:prob_fact}
\end{multline}
with all contours around the origin. The fact that the integrand has a factorised form is non-trivial, but this form makes it amenable to asymptotic analysis. This type of multiple integral formula with two sets of variables have appeared in a few different contexts \cite{KMMMP1993, Kostov1996}. 
From now on we take  $\a=\b=\frac12$ for simplicity. 

Without derivation we note that it is possible to write \eqref{eq:prob_fact} as a single determinant of the form
\begin{align}
P_{N,M,\rho}(t) = c_{NM}(t) \det \Big( \sum_{x\ge 0} f_j(x)g_k(x) \Big)_{j,k=1}^M,
\label{singledet}
\end{align}
where $c_{NM}(t) = \e^{-(N+M) t}  2^{NM} \rho^N$ and
\begin{align*}
f_j(x) &= \oint \frac{\dd w}{2\pi \ii}  \e^{wt/2} \frac{w^{x +j-1}} {(w-1)^j},\qquad 1\le j \le M,\\
g_k(x) &= \oint  \frac{\dd z }{2\pi \ii} \e^{zt/2} \frac{(-z)^{-x+k-2}}{(z-1)^{k}(1-(1-\rho)z)},\ 1\le k \le N,
\end{align*}
where the integrals are again around the origin and $g_k(x) = \delta_{x,M-k}$ for $N < k \leq M$. 
The geometric sum over $x$ in \eqref{singledet} can be easily performed. After changing the contours to lie around the other poles, the contour integrals for $f$ and $g$ can be computed and a single determinant remains with explicitly known matrix elements. 
This explicit representation resulting from \eqref{singledet} is useful for numerical evaluation.

\paragraph{Nonlinear fluctuating hydrodynamics.} We now summarise the predictions from the general theory of NLFHD 
\cite{Spohn2014} applied to the AHR model \cite{FSS2013}, and show below how these are confirmed by a precise asymptotic analysis of \eqref{eq:prob_fact}.  In fact, the AHR model has served as a prototypical model for checking numerically predictions of NLFHD. The case of a step initial condition was studied in \cite{MS2016,MS2017}. 

First we describe the predictions for the case in which infinitely many $+$ particles are Bernoulli distributed with density $\rho$ on the negative integers, and infinitely many $-$ particles fill the lattice completely on the non-negative integers. The macroscopic behavior of this system at the Euler scale can be described by a hydrodynamic equation of the form
\cite{FT2004},
\begin{align}
	\frac{\partial \boldsymbol u(x,t)}{\partial t}+\frac{\partial\boldsymbol{\mathsf j}(\boldsymbol u(x,t))}{\partial x}=0,
\end{align}
where $\boldsymbol u(x,t) = (\rho_+(x,t),\rho_-(x,t))$ is the density vector and
$\boldsymbol{\mathsf j}(\boldsymbol u) = (\mathsf j_{+}(\boldsymbol u), \mathsf j_{-}(\boldsymbol u))$ denotes the macroscopic current of $\pm$ particles given by 
\begin{align}
	&\mathsf j_{+} (\boldsymbol u)=\rho_+(1-\rho_+-\rho_-) + 2\rho_+\rho_-,\label{currenth}\\
	&\mathsf j_{-} (\boldsymbol u)=-(1-\rho_+-\rho_-)\rho_- -2\rho_+\rho_- \ . \label{currentc}
\end{align}
This set of coupled equations with step type initial condition (Riemann problem) can be solved by switching to the 
normal modes that diagonalize the Jacobian $\partial \boldsymbol{\mathsf j}/ \partial \boldsymbol{\mathsf u}$
\cite{Bressan2009,MS2017}. 
The average currents of $\pm$ particles at the origin are given by $j_+=\rho(3-\rho)^2/16$, and $j_-=(1+\rho)^2(2-\rho)/16$, where we recall that $\rho$ is the initial average density of $+$ particles. We are interested in the fluctuations around 
these values.

In fluctuating hydrodynamics, one presumes that the fluctuations of the model can be taken into account by adding noise and diffusion terms to the hydrodynamic equations \cite{Spohn2014}. For the AHR model, the equations for the two modes become a coupled KPZ equation 
\cite{EK1992, FSS2013,FH2016}. Because the speeds of the two normal modes are different, one can naively expect that, in the long time limit, the fluctuations of each mode is described by the KPZ equation. This is the basic idea behind NLFHD. 

For a system with infinitely many particles, NLFHD predicts that the probability of observing
$n_+(t)=n$ and $n_-(t)=m$ in the long time scaling limit is approximately given by    
\begin{equation}
\mathbb{P}_{\infty,\infty}[n_+(t)=n,n_-(t)=m] \simeq F'_G(s_{+1})F'_2(s_{-1})
\label{conj}
\end{equation}
where $F_G$ and $F_2$ are the Gaussian and GUE Tracy-Widom distributions respectively,  
and the two scaling variables $s_{-}=s_{-}(n,m;t)$ and $s_{+}=s_{+}(n,m;t)$ 
associated with the two normal modes are given by
\begin{multline}
s_{-}=  \Big( (1+\rho) n-(3-\rho)m+(1-\rho) (1-\frac{(1-\rho)^2}{4}) t\Big)/ \\
\Big( (3/16)^{1/3}(1-\rho)(3-\rho)^{2/3}(1+\rho)^{2/3}t^{1/3} \Big),
\label{sm}
\end{multline}
\vskip-20pt
\begin{multline}
s_{+}=
\Big(  -2(2-\rho) n+2\rho m+2(2-\rho)(1-\rho)\rho t \Big)/ \\
\Big(3(1-\rho)^{3/2}\sqrt{\rho(2-\rho)}t^{1/2} \Big).
\label{sp}
\end{multline}
The product structure of the distribution in (\ref{conj}) is implied by the anticipated independence of the two normal modes. This fact is naturally expected in NLFHD but to our knowledge has not been explicitly stated before.   

In \eqref{eq:prob_fact} we found an exact formula for the quantity $\mathbb{P}_{N,M}[n_+(t)=N,n_-(t)=M]$, the probability of observing $n_+(t)=N$ and $n_-(t)=M$ given an initial condition of a \emph{finite} 
number $N$ Bernoulli distributed $+$ particles at density $\rho$ to the left of the origin, and $M$ 
fully packed $-$ particles to the right of the origin. One immediately notices however that $\mathbb{P}_{N,M}[n_+(t)=N,n_-(t)=M]$ is very different from $\mathbb{P}_{\infty,\infty}[n_+(t)=N,n_-(t)=M]$. For example,  when $t\to\infty$ for fixed $N,M$,  
the latter tends to zero whilst the former approaches unity. 

It is however possible to generalize predictions of NLFHD for the AHR model to the case with finite number of particles. The idea is to make a connection between the probability for the system with finite number of particles to a similar probability for the system with infinite number of particles on the scale where $s_\pm$ as defined in \eqref{sm} and \eqref{sp} are well defined. 

\paragraph{Prediction for finite number of particles.} Our third main result is a prediction resulting from NLFHD for systems with large but finite number of particles. Let us consider the probability $\mathbb{P}_{N,M}[n_+(t)=n,n_-(t)=m]$, for the case where $N$ and $M$ are chosen in such a way that $+$ and $-$ particles cross the origin about the same time $t$, as in FIG. \ref{fig:AHRicex}, and when we consider the fluctuations around this time. For $n<N$ and $m<M$, this is the same as $\mathbb{P}_{\infty,\infty}[n_+(t)=n,n_-(t)=m]$. For the case where all $+$ particles have crossed the origin but not yet all $-$ particles, i.e. $n=N$ and $m<M$, 
\be
\mathbb{P}_{N,M}[n_+(t)=N,n_-(t)=m]  \simeq \sum_{s' \ge \tilde{s}_-}  F(\tilde{s}_+,s'),
\ee
where $F(s_+,s_-):=\mathbb{P}_{\infty,\infty}[n_+(t)=n,n_-(t)=m]$ and $\tilde{s}_\pm = s_\pm (N,m,t)$. A similar argument applies to the case with $n<N$ and $m=M$.
The probability in the case of $n=N$ and $m=M$ is approximated by the remaining sum of probabilities for infinitely many particles, i.e. 
\be 
\mathbb{P}_{N,M}[n_+(t)=N,n_-(t)=M] \simeq \sum_{s\ge s_+^*} \sum_{s'\ge s_-^*} F(s,s'),
\ee 
where $s_\pm^* = s_\pm(N,M,t)$.
In the scaling regime defined by $s_\pm$, using \eqref{conj} this simply becomes, 
\begin{equation}
\mathbb{P}_{N, M}[n_+(t)=N,n_-(t)=M] \simeq F_G(s_{+}^*)F_2(s_{-}^*). 
\label{conjf}
\end{equation}

\begin{figure}
\scalebox{0.3}{\includegraphics{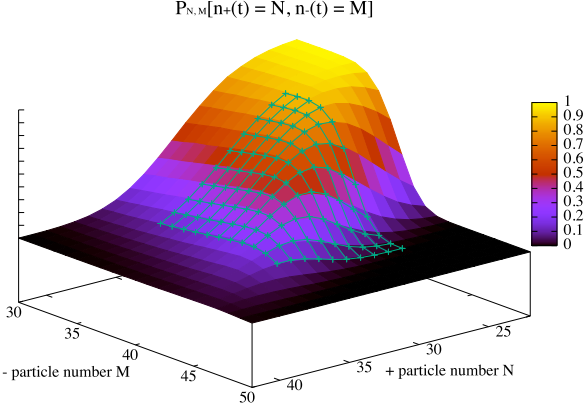}}
\caption{A 3D plot of the probability $P_{N,M,\rho}(t)$ from Monte Carlo simulations (gradual colors) and 
the conjecture (\ref{conjf}) (green mesh).}
\label{fig:3d}
\end{figure}

Equation (\ref{conjf}) can be considered as the prediction of NLFHD for the case with finite particles. We emphasize that this is a nontrivial generalization of predictions of NLFHD for the case with finite number of particles. In FIG. \ref{fig:3d} we show the 3D plot of the probabilities from Monte Carlo simulations and the conjecture (\ref{conjf}). A good agreement is observed and it gives an evidence that our generalized conjecture of NLFHD (\ref{conjf}) is indeed true. 

\paragraph{Asymptotic analysis for $N\geq M$.}
The case of $N\geq M$ is simplest to analyse though it does not correspond to the correct scaling regime 
for (\ref{sm}),(\ref{sp}) as $-$ particles will have crossed the origin long before all $+$ particles have. In this regime there is no pole at $w_k=\infty$ and the $w$ integration in \eqref{eq:prob_fact} can be performed successively by deforming the contour to lie around the only other pole at $w_k=1$. The remaining $N$-fold integral, using known methods \cite{IS2017p}, 
can be rewritten as a Fredholm determinant, giving
\be
P_{N,M,\rho}(t) = \det(1-K(\xi,\eta))_{\ell_2(\mathbb{N})},
\label{fredholm}
\ee
with kernel $K(\xi,\eta) = \sum_{j=0}^{N-1} \phi_k(\xi) \psi_k(\eta)$, and
\begin{align}
 \phi_k(\xi) &= \int_D \frac{\dd w}{2\pi\ii} \frac{w^{k-\xi}(1+w)^M(1-(1-\rho)/w)}{w^M(w-1)^{k+1}e^{wt/2}},
 \label{phi} \\
 \psi_k(\eta) &= \int_{C} \frac{\dd z}{2\pi\ii} \frac{z^M (z-1)^k e^{zt/2}}{z^{k+2-\eta}(1+z)^M(1-(1-\rho)/z)},
 \label{psi}
\end{align}
for $1\leq k\leq N$ 
where $D$ is a contour around 1 and $C$ encloses $0,-1,1-\rho$. 
Standard asymptotic analysis \cite{Johansson2000, BFP2007} then shows that the limit is governed by the 
GUE Tracy-Widom distribution, as expected since $P_{N,M,\rho}(t)$ in this region should asymptotically be
close to the current distribution of the single species TASEP \cite{Johansson2000,PS2002a}. 

\paragraph{Asymptotic analysis for $N<M$.}
Here we give our fourth main result, an analytic confirmation of (\ref{conjf}) by performing asymptotic analysis to our exact formula (\ref{eq:prob_fact}). We first deform the $z$-contours in \eqref{eq:prob_fact} to lie around all other poles in the $z$-plane other than the origin. As there are no poles at $z_j=\infty$ when $N<M$, the only remaining poles are at $z_j=1$ and $z_j=1/(1-\rho)$. 
By evaluating the simple pole at $z_j=1/(1-\rho)$ in (\ref{eq:prob_fact}), we obtain 
\begin{equation}
P_{N,M,\rho}(t)
=
I_1 + J \times I_2,
\label{eq:IJI}
\end{equation} 
where 
\begin{align}
I_1
&=
\frac{1}{M!} \oint \prod_{k=1}^M \frac{\dd w_k}{2\pi\ii }  \frac{\e^{\Lambda_{0,M} t}\Delta_M(w)} {\prod_{k=1}^M (\frac12(1+w_k))^N} ,\notag\\
J &= \frac{\rho^{N-1}}{(1-\rho)^N}\left(\frac{2(1-\rho)}{2-\rho}\right)^M \frac{\e^{-\rho t/2}}{(N-1)!} \oint \prod_{j=1}^{N-1} 
\frac{\dd z_j}{2\pi\ii} 
\notag\\
&~~\times 
\e^{\Lambda_{N-1,0}t} \frac{\Delta_{N-1}(z)\prod_{j=1}^{N-1}(1-(1-\rho)z_j)}  {\prod_{j=1}^{N-1} (z_j-1) \prod_{j=1}^{N-1} (\frac12(1+z_j))^M}, 
\notag\\
I_2
&=
\frac{1}{M!} \oint\prod_{k=1}^M \frac{\dd w_k}{2\pi\ii } \e^{\Lambda_{0,M}t}\Delta_M(w)
\notag\\   
&\quad\times   
\frac{\prod_{j=1}^{N-1} (1+z_j)^M (\frac{1}{1-\rho}+1)^M}
       {\prod_{k=1}^M(\prod_{j=1}^{N-1}(z_j+w_k)(\frac{1}{1-\rho}+w_k))} .        
\end{align} 
Here $\Delta_N(z) = \prod_{1\le i<j \le N} (z_i-z_j)^2/\prod_{j=1}^N (z_j-1)^N$, 
the $w$-integration is around the origin and the $z$-integration is around $z_j=1$ only.
It can be shown \cite{ChenGHS} that the leading order contribution to the asymptotics of $I_2$ is given by $z_j=1$. Hence we can ignore the $z$-dependence in $I_2$ by substituting $z_j=1$ and then evaluate the asymptotic behaviour of $I_2$ and $J$ independently.

Performing an asymptotic analysis similar to that in \cite{IS2017p}, the integrals $I_{1,2}$ can be treated analogously to the case 
$N\geq M$, with formulas similar to \eqref{fredholm}-\eqref{psi}, 
resulting in GUE Tracy-Widom distributions, i.e. $I_{1,2}\simeq F_2(s_-)$. The integral $J$ can also be analysed using similar methods and one finds $J\simeq F_G(s_+)-1$. In light of \eqref{eq:IJI} this establishes our conjecture (\ref{conjf}) of NLFHD.
By estimating lower order contributions around $z_j=1$ in $I_2$, we would be able to study the interaction effects between the 
modes, but this is left as an important problem for future research. 


To conclude, in this Letter we presented the exact Green's function for the two-species AHR exclusion process for 
certain initial and final conditions, and provided an exact multiple integral formula for a joint current distribution. 
For a mix of step and Bernoulli type initial conditions, we have performed an asymptotic analysis for this multiple 
integral in the scaling limit. We found that the limiting distribution is given by a product of the Gaussian and the 
GUE Tracy-Widom distribution, as predicted by a generalization of NLFHD for finite number of particles. 
This is the first analytic confirmation of predictions of 1D non-linear fluctuating hydrodynamics for multi-component systems.

Due to the fact that we have explicit knowledge of the full Green's function for the AHR model, our approach can be generalized to study various other initial conditions and other observables. It is also our firm expectation that our approach should generalize to other integrable multi-species models, such as the ones studied in \cite{TW2013, Kuan2018p}. Indeed, our work opens up new ways to analytically study fluctuations and correlation properties of many multi-component systems.


\begin{thebibliography}{78}%
\makeatletter
\providecommand \@ifxundefined [1]{%
 \@ifx{#1\undefined}
}%
\providecommand \@ifnum [1]{%
 \ifnum #1\expandafter \@firstoftwo
 \else \expandafter \@secondoftwo
 \fi
}%
\providecommand \@ifx [1]{%
 \ifx #1\expandafter \@firstoftwo
 \else \expandafter \@secondoftwo
 \fi
}%
\providecommand \natexlab [1]{#1}%
\providecommand \enquote  [1]{``#1''}%
\providecommand \bibnamefont  [1]{#1}%
\providecommand \bibfnamefont [1]{#1}%
\providecommand \citenamefont [1]{#1}%
\providecommand \href@noop [0]{\@secondoftwo}%
\providecommand \href [0]{\begingroup \@sanitize@url \@href}%
\providecommand \@href[1]{\@@startlink{#1}\@@href}%
\providecommand \@@href[1]{\endgroup#1\@@endlink}%
\providecommand \@sanitize@url [0]{\catcode `\\12\catcode `\$12\catcode
  `\&12\catcode `\#12\catcode `\^12\catcode `\_12\catcode `\%12\relax}%
\providecommand \@@startlink[1]{}%
\providecommand \@@endlink[0]{}%
\providecommand \url  [0]{\begingroup\@sanitize@url \@url }%
\providecommand \@url [1]{\endgroup\@href {#1}{\urlprefix }}%
\providecommand \urlprefix  [0]{URL }%
\providecommand \Eprint [0]{\href }%
\providecommand \doibase [0]{http://dx.doi.org/}%
\providecommand \selectlanguage [0]{\@gobble}%
\providecommand \bibinfo  [0]{\@secondoftwo}%
\providecommand \bibfield  [0]{\@secondoftwo}%
\providecommand \translation [1]{[#1]}%
\providecommand \BibitemOpen [0]{}%
\providecommand \bibitemStop [0]{}%
\providecommand \bibitemNoStop [0]{.\EOS\space}%
\providecommand \EOS [0]{\spacefactor3000\relax}%
\providecommand \BibitemShut  [1]{\csname bibitem#1\endcsname}%
\let\auto@bib@innerbib\@empty
\bibitem [{\citenamefont {{L. D. Landau, E. M. Lifshitz}}(1959)}]{LanLif}%
  \BibitemOpen
  \bibfield  {author} {\bibinfo {author} {\bibnamefont {{L. D. Landau, E. M.
  Lifshitz}}},\ }\href@noop {} {\emph {\bibinfo {title} {Fluid Mechanics, Ch.
  17}}}\ (\bibinfo  {publisher} {Pergamon Press},\ \bibinfo {year}
  {1959})\BibitemShut {NoStop}%
\bibitem [{\citenamefont {{S. Lepri, R. Livi, A. Politi}}(2003)}]{LLP2003}%
  \BibitemOpen
  \bibfield  {author} {\bibinfo {author} {\bibnamefont {{S. Lepri, R. Livi, A.
  Politi}}},\ }\href@noop {} {\bibfield  {journal} {\bibinfo  {journal} {Phys.
  Rep.}\ }\textbf {\bibinfo {volume} {37}},\ \bibinfo {pages} {1} (\bibinfo
  {year} {2003})}\BibitemShut {NoStop}%
\bibitem [{\citenamefont {{G. Basile, C. Bernardin, and S.
  Olla}}(2006)}]{BBO2006}%
  \BibitemOpen
  \bibfield  {author} {\bibinfo {author} {\bibnamefont {{G. Basile, C.
  Bernardin, and S. Olla}}},\ }\href@noop {} {\bibfield  {journal} {\bibinfo
  {journal} {Phys. Rev. Lett.}\ }\textbf {\bibinfo {volume} {96}},\ \bibinfo
  {pages} {204303} (\bibinfo {year} {2006})}\BibitemShut {NoStop}%
\bibitem [{\citenamefont {{A. Dhar}}(2008)}]{Dhar2008}%
  \BibitemOpen
  \bibfield  {author} {\bibinfo {author} {\bibnamefont {{A. Dhar}}},\
  }\href@noop {} {\bibfield  {journal} {\bibinfo  {journal} {Adv. Phys.}\
  }\textbf {\bibinfo {volume} {57}},\ \bibinfo {pages} {457} (\bibinfo {year}
  {2008})}\BibitemShut {NoStop}%
\bibitem [{\citenamefont {Lepri}(2016)}]{Lepri2016}%
  \BibitemOpen
  \bibinfo {editor} {\bibfnamefont {S.}~\bibnamefont {Lepri}},\ ed.,\
  \href@noop {} {\emph {\bibinfo {title} {Thermal Transport in Low
  Dimensions}}}\ (\bibinfo  {publisher} {Springer-Verlag},\ \bibinfo {year}
  {2016})\BibitemShut {NoStop}%
\bibitem [{\citenamefont {{H. Mori, H. Fujisaka}}(1973)}]{MF1973}%
  \BibitemOpen
  \bibfield  {author} {\bibinfo {author} {\bibnamefont {{H. Mori, H.
  Fujisaka}}},\ }\href@noop {} {\bibfield  {journal} {\bibinfo  {journal}
  {Prog. Theo. Phys.}\ }\textbf {\bibinfo {volume} {49}},\ \bibinfo {pages}
  {764} (\bibinfo {year} {1973})}\BibitemShut {NoStop}%
\bibitem [{\citenamefont {{J. Swift and P.C. Hohenberg}}(1977)}]{SH1977}%
  \BibitemOpen
  \bibfield  {author} {\bibinfo {author} {\bibnamefont {{J. Swift and P.C.
  Hohenberg}}},\ }\href@noop {} {\bibfield  {journal} {\bibinfo  {journal}
  {Phys. Rev. A}\ }\textbf {\bibinfo {volume} {15}},\ \bibinfo {pages} {319}
  (\bibinfo {year} {1977})}\BibitemShut {NoStop}%
\bibitem [{\citenamefont {{S. P. Das, G. F. Mazenko}}(1986)}]{DasMazenko1986}%
  \BibitemOpen
  \bibfield  {author} {\bibinfo {author} {\bibnamefont {{S. P. Das, G. F.
  Mazenko}}},\ }\href@noop {} {\bibfield  {journal} {\bibinfo  {journal} {Phys.
  Rev. A.}\ }\textbf {\bibinfo {volume} {34}},\ \bibinfo {pages} {2265}
  (\bibinfo {year} {1986})}\BibitemShut {NoStop}%
\bibitem [{\citenamefont {{M. Kardar and G. Parisi and Y. C.
  Zhang}}(1986)}]{KPZ1986}%
  \BibitemOpen
  \bibfield  {author} {\bibinfo {author} {\bibnamefont {{M. Kardar and G.
  Parisi and Y. C. Zhang}}},\ }\href@noop {} {\bibfield  {journal} {\bibinfo
  {journal} {Phys. Rev. Lett.}\ }\textbf {\bibinfo {volume} {56}},\ \bibinfo
  {pages} {889} (\bibinfo {year} {1986})}\BibitemShut {NoStop}%
\bibitem [{\citenamefont {{A.L. Barab{\'a}si and H.E.
  Stanley}}(1995)}]{BS1995}%
  \BibitemOpen
  \bibfield  {author} {\bibinfo {author} {\bibnamefont {{A.L. Barab{\'a}si and
  H.E. Stanley}}},\ }\href@noop {} {\emph {\bibinfo {title} {Fractal concepts
  in surface growth}}}\ (\bibinfo  {publisher} {Cambridge University Press},\
  \bibinfo {year} {1995})\BibitemShut {NoStop}%
\bibitem [{\citenamefont {van Beijeren}(2012)}]{Beijeren2012}%
  \BibitemOpen
  \bibfield  {author} {\bibinfo {author} {\bibfnamefont {H.}~\bibnamefont {van
  Beijeren}},\ }\href@noop {} {\bibfield  {journal} {\bibinfo  {journal} {Phys.
  Rev. Lett.}\ }\textbf {\bibinfo {volume} {108}},\ \bibinfo {pages} {180601}
  (\bibinfo {year} {2012})}\BibitemShut {NoStop}%
\bibitem [{\citenamefont {Pr{\"a}hofer}\ and\ \citenamefont
  {Spohn}(2004)}]{PS2004}%
  \BibitemOpen
  \bibfield  {author} {\bibinfo {author} {\bibfnamefont {M.}~\bibnamefont
  {Pr{\"a}hofer}}\ and\ \bibinfo {author} {\bibfnamefont {H.}~\bibnamefont
  {Spohn}},\ }\href@noop {} {\bibfield  {journal} {\bibinfo  {journal} {J.
  Stat. Phys.}\ }\textbf {\bibinfo {volume} {115}},\ \bibinfo {pages} {255}
  (\bibinfo {year} {2004})}\BibitemShut {NoStop}%
\bibitem [{\citenamefont {Ferrari}\ and\ \citenamefont {Spohn}(2006)}]{FS2006}%
  \BibitemOpen
  \bibfield  {author} {\bibinfo {author} {\bibfnamefont {P.~L.}\ \bibnamefont
  {Ferrari}}\ and\ \bibinfo {author} {\bibfnamefont {H.}~\bibnamefont
  {Spohn}},\ }\href@noop {} {\bibfield  {journal} {\bibinfo  {journal} {Comm.
  Math. Phys.}\ }\textbf {\bibinfo {volume} {265}},\ \bibinfo {pages} {1}
  (\bibinfo {year} {2006})}\BibitemShut {NoStop}%
\bibitem [{\citenamefont {Imamura}\ and\ \citenamefont
  {Sasamoto}(2012)}]{IS2012}%
  \BibitemOpen
  \bibfield  {author} {\bibinfo {author} {\bibfnamefont {T.}~\bibnamefont
  {Imamura}}\ and\ \bibinfo {author} {\bibfnamefont {T.}~\bibnamefont
  {Sasamoto}},\ }\href@noop {} {\bibfield  {journal} {\bibinfo  {journal}
  {Phys. Rev. Lett.}\ }\textbf {\bibinfo {volume} {108}},\ \bibinfo {pages}
  {190603} (\bibinfo {year} {2012})}\BibitemShut {NoStop}%
\bibitem [{\citenamefont {Imamura}\ and\ \citenamefont
  {Sasamoto}(2013)}]{IS2013}%
  \BibitemOpen
  \bibfield  {author} {\bibinfo {author} {\bibfnamefont {T.}~\bibnamefont
  {Imamura}}\ and\ \bibinfo {author} {\bibfnamefont {T.}~\bibnamefont
  {Sasamoto}},\ }\href@noop {} {\bibfield  {journal} {\bibinfo  {journal} {J.
  Stat. Phys.}\ }\textbf {\bibinfo {volume} {150}},\ \bibinfo {pages} {908}
  (\bibinfo {year} {2013})}\BibitemShut {NoStop}%
\bibitem [{\citenamefont {Spohn}(2014)}]{Spohn2014}%
  \BibitemOpen
  \bibfield  {author} {\bibinfo {author} {\bibfnamefont {H.}~\bibnamefont
  {Spohn}},\ }\href@noop {} {\bibfield  {journal} {\bibinfo  {journal} {J.
  Stat. Phys.}\ }\textbf {\bibinfo {volume} {154}},\ \bibinfo {pages} {1191}
  (\bibinfo {year} {2014})}\BibitemShut {NoStop}%
\bibitem [{\citenamefont {{E. Fermi, J. Pasta, S. Ulam}}(1955)}]{FPU1955}%
  \BibitemOpen
  \bibfield  {author} {\bibinfo {author} {\bibnamefont {{E. Fermi, J. Pasta, S.
  Ulam}}},\ }\href@noop {} {\bibfield  {journal} {\bibinfo  {journal} {Los
  Alamos Report LA-1940}\ } (\bibinfo {year} {1955})}\BibitemShut {NoStop}%
\bibitem [{\citenamefont {{G. Gallavotti}}(2008)}]{Gallavotti2008}%
  \BibitemOpen
  \bibfield  {author} {\bibinfo {author} {\bibnamefont {{G. Gallavotti}}},\
  }\href@noop {} {\emph {\bibinfo {title} {{The Fermi-Pasta-Ulam Problem: A
  Status Report. Lecture Notes in Physics 728}}}}\ (\bibinfo  {publisher}
  {Springer, Berlin},\ \bibinfo {year} {2008})\BibitemShut {NoStop}%
\bibitem [{\citenamefont {{P. Ferrari, T. Sasamoto and H.
  Spohn}}(2013)}]{FSS2013}%
  \BibitemOpen
  \bibfield  {author} {\bibinfo {author} {\bibnamefont {{P. Ferrari, T.
  Sasamoto and H. Spohn}}},\ }\href@noop {} {\bibfield  {journal} {\bibinfo
  {journal} {J. Stat. Phys.}\ }\textbf {\bibinfo {volume} {153}},\ \bibinfo
  {pages} {377} (\bibinfo {year} {2013})}\BibitemShut {NoStop}%
\bibitem [{\citenamefont {{C. B. Mendl, H. Spohn}}(2016)}]{MS2016}%
  \BibitemOpen
  \bibfield  {author} {\bibinfo {author} {\bibnamefont {{C. B. Mendl, H.
  Spohn}}},\ }\href@noop {} {\bibfield  {journal} {\bibinfo  {journal} {Phys.
  Rev. E.}\ }\textbf {\bibinfo {volume} {93}},\ \bibinfo {pages} {060101(R)}
  (\bibinfo {year} {2016})}\BibitemShut {NoStop}%
\bibitem [{\citenamefont {{C. B. Mendl, H. Spohn}}(2017)}]{MS2017}%
  \BibitemOpen
  \bibfield  {author} {\bibinfo {author} {\bibnamefont {{C. B. Mendl, H.
  Spohn}}},\ }\href@noop {} {\bibfield  {journal} {\bibinfo  {journal} {J.
  Stat. Phys.}\ }\textbf {\bibinfo {volume} {166}},\ \bibinfo {pages} {841}
  (\bibinfo {year} {2017})}\BibitemShut {NoStop}%
\bibitem [{\citenamefont {{S. G. Das, A. Dhar, K. Saito, C. B. Mendl, H.
  Spohn}}(2014)}]{DDSMS2014}%
  \BibitemOpen
  \bibfield  {author} {\bibinfo {author} {\bibnamefont {{S. G. Das, A. Dhar, K.
  Saito, C. B. Mendl, H. Spohn}}},\ }\href@noop {} {\bibfield  {journal}
  {\bibinfo  {journal} {Phys. Rev. E.}\ }\textbf {\bibinfo {volume} {90}},\
  \bibinfo {pages} {012124} (\bibinfo {year} {2014})}\BibitemShut {NoStop}%
\bibitem [{\citenamefont {{C. B. Mendl, H. Spohn}}(2013)}]{MS2013}%
  \BibitemOpen
  \bibfield  {author} {\bibinfo {author} {\bibnamefont {{C. B. Mendl, H.
  Spohn}}},\ }\href@noop {} {\bibfield  {journal} {\bibinfo  {journal} {Phys.
  Rev. Lett.}\ }\textbf {\bibinfo {volume} {111}},\ \bibinfo {pages} {230601}
  (\bibinfo {year} {2013})}\BibitemShut {NoStop}%
\bibitem [{\citenamefont {{C. B. Mendl, H. Spohn}}(2014)}]{MS2014}%
  \BibitemOpen
  \bibfield  {author} {\bibinfo {author} {\bibnamefont {{C. B. Mendl, H.
  Spohn}}},\ }\href@noop {} {\bibfield  {journal} {\bibinfo  {journal} {Phys.
  Rev. E.}\ }\textbf {\bibinfo {volume} {90}},\ \bibinfo {pages} {012147}
  (\bibinfo {year} {2014})}\BibitemShut {NoStop}%
\bibitem [{\citenamefont {{C. B. Mendl, H. Spohn}}(2015)}]{MS2015}%
  \BibitemOpen
  \bibfield  {author} {\bibinfo {author} {\bibnamefont {{C. B. Mendl, H.
  Spohn}}},\ }\href@noop {} {\bibfield  {journal} {\bibinfo  {journal} {J.
  Stat. Mech.}\ }\textbf {\bibinfo {volume} {2015}},\ \bibinfo {pages} {P03007}
  (\bibinfo {year} {2015})}\BibitemShut {NoStop}%
\bibitem [{\citenamefont {{S. Lepri, H. Bufferand, G. Ciraolo, P. Di Cintio, P.
  Ghendrih, R. Livi}}()}]{LBCCGL2018p}%
  \BibitemOpen
  \bibfield  {author} {\bibinfo {author} {\bibnamefont {{S. Lepri, H.
  Bufferand, G. Ciraolo, P. Di Cintio, P. Ghendrih, R. Livi}}},\ }\href@noop {}
  {\ }\Eprint {http://arxiv.org/abs/arXiv:1801.09944} {arXiv:1801.09944}
  \BibitemShut {NoStop}%
\bibitem [{\citenamefont {{M. Kulkarni, D. A. Huse, H.
  Spohn}}(2015)}]{KHS2015}%
  \BibitemOpen
  \bibfield  {author} {\bibinfo {author} {\bibnamefont {{M. Kulkarni, D. A.
  Huse, H. Spohn}}},\ }\href@noop {} {\bibfield  {journal} {\bibinfo  {journal}
  {Phys. Rev. A}\ }\textbf {\bibinfo {volume} {92}},\ \bibinfo {pages} {043612}
  (\bibinfo {year} {2015})}\BibitemShut {NoStop}%
\bibitem [{\citenamefont {{V. Popkov, J. Schmidt, G. M.
  Sch\"utz}}(2014)}]{PSS2014}%
  \BibitemOpen
  \bibfield  {author} {\bibinfo {author} {\bibnamefont {{V. Popkov, J. Schmidt,
  G. M. Sch\"utz}}},\ }\href@noop {} {\bibfield  {journal} {\bibinfo  {journal}
  {Phys. Rev. Lett.}\ }\textbf {\bibinfo {volume} {112}},\ \bibinfo {pages}
  {200602} (\bibinfo {year} {2014})}\BibitemShut {NoStop}%
\bibitem [{\citenamefont {{V. Popkov, J. Schmidt, G. M.
  Sch\"utz}}(2015)}]{PSS2015}%
  \BibitemOpen
  \bibfield  {author} {\bibinfo {author} {\bibnamefont {{V. Popkov, J. Schmidt,
  G. M. Sch\"utz}}},\ }\href@noop {} {\bibfield  {journal} {\bibinfo  {journal}
  {J. Stat. Phys.}\ }\textbf {\bibinfo {volume} {160}},\ \bibinfo {pages} {835}
  (\bibinfo {year} {2015})}\BibitemShut {NoStop}%
\bibitem [{\citenamefont {Corwin}(2012)}]{Corwin2012}%
  \BibitemOpen
  \bibfield  {author} {\bibinfo {author} {\bibfnamefont {I.}~\bibnamefont
  {Corwin}},\ }\href@noop {} {\bibfield  {journal} {\bibinfo  {journal} {Random
  Matrices Theory Appl.}\ }\textbf {\bibinfo {volume} {1}} (\bibinfo {year}
  {2012})}\BibitemShut {NoStop}%
\bibitem [{\citenamefont {{J. Quastel and H. Spohn}}(2015)}]{QS2015}%
  \BibitemOpen
  \bibfield  {author} {\bibinfo {author} {\bibnamefont {{J. Quastel and H.
  Spohn}}},\ }\href@noop {} {\bibfield  {journal} {\bibinfo  {journal} {J.
  Stat. Phys.}\ }\textbf {\bibinfo {volume} {160}},\ \bibinfo {pages} {965}
  (\bibinfo {year} {2015})}\BibitemShut {NoStop}%
\bibitem [{\citenamefont {Johansson}(2000)}]{Johansson2000}%
  \BibitemOpen
  \bibfield  {author} {\bibinfo {author} {\bibfnamefont {K.}~\bibnamefont
  {Johansson}},\ }\href@noop {} {\bibfield  {journal} {\bibinfo  {journal}
  {Comm. Math. Phys.}\ }\textbf {\bibinfo {volume} {209}},\ \bibinfo {pages}
  {437} (\bibinfo {year} {2000})}\BibitemShut {NoStop}%
\bibitem [{\citenamefont {Tracy}\ and\ \citenamefont {Widom}(2009)}]{TW2009a}%
  \BibitemOpen
  \bibfield  {author} {\bibinfo {author} {\bibfnamefont {C.~A.}\ \bibnamefont
  {Tracy}}\ and\ \bibinfo {author} {\bibfnamefont {H.}~\bibnamefont {Widom}},\
  }\href@noop {} {\bibfield  {journal} {\bibinfo  {journal} {Comm. Math.
  Phys.}\ }\textbf {\bibinfo {volume} {209}},\ \bibinfo {pages} {129} (\bibinfo
  {year} {2009})}\BibitemShut {NoStop}%
\bibitem [{\citenamefont {Sasamoto}\ and\ \citenamefont
  {Spohn}(2010{\natexlab{a}})}]{SS2010a}%
  \BibitemOpen
  \bibfield  {author} {\bibinfo {author} {\bibfnamefont {T.}~\bibnamefont
  {Sasamoto}}\ and\ \bibinfo {author} {\bibfnamefont {H.}~\bibnamefont
  {Spohn}},\ }\href@noop {} {\bibfield  {journal} {\bibinfo  {journal} {J.
  Stat. Phys.}\ }\textbf {\bibinfo {volume} {140}},\ \bibinfo {pages} {209}
  (\bibinfo {year} {2010}{\natexlab{a}})}\BibitemShut {NoStop}%
\bibitem [{\citenamefont {Sasamoto}\ and\ \citenamefont
  {Spohn}(2010{\natexlab{b}})}]{SS2010b}%
  \BibitemOpen
  \bibfield  {author} {\bibinfo {author} {\bibfnamefont {T.}~\bibnamefont
  {Sasamoto}}\ and\ \bibinfo {author} {\bibfnamefont {H.}~\bibnamefont
  {Spohn}},\ }\href@noop {} {\bibfield  {journal} {\bibinfo  {journal} {Nucl.
  Phys. B}\ }\textbf {\bibinfo {volume} {834}},\ \bibinfo {pages} {523}
  (\bibinfo {year} {2010}{\natexlab{b}})}\BibitemShut {NoStop}%
\bibitem [{\citenamefont {Sasamoto}\ and\ \citenamefont
  {Spohn}(2010{\natexlab{c}})}]{SS2010c}%
  \BibitemOpen
  \bibfield  {author} {\bibinfo {author} {\bibfnamefont {T.}~\bibnamefont
  {Sasamoto}}\ and\ \bibinfo {author} {\bibfnamefont {H.}~\bibnamefont
  {Spohn}},\ }\href@noop {} {\bibfield  {journal} {\bibinfo  {journal} {Phys.
  Rev. Lett.}\ }\textbf {\bibinfo {volume} {104}},\ \bibinfo {pages} {230602}
  (\bibinfo {year} {2010}{\natexlab{c}})}\BibitemShut {NoStop}%
\bibitem [{\citenamefont {{G. Amir and I. Corwin and J.
  Quastel}}(2011)}]{ACQ2011}%
  \BibitemOpen
  \bibfield  {author} {\bibinfo {author} {\bibnamefont {{G. Amir and I. Corwin
  and J. Quastel}}},\ }\href@noop {} {\bibfield  {journal} {\bibinfo  {journal}
  {Comm. Pure Appl. Math.}\ }\textbf {\bibinfo {volume} {64}},\ \bibinfo
  {pages} {466} (\bibinfo {year} {2011})}\BibitemShut {NoStop}%
\bibitem [{\citenamefont {Baik}\ and\ \citenamefont {Rains}(2000)}]{BR2000}%
  \BibitemOpen
  \bibfield  {author} {\bibinfo {author} {\bibfnamefont {J.}~\bibnamefont
  {Baik}}\ and\ \bibinfo {author} {\bibfnamefont {E.~M.}\ \bibnamefont
  {Rains}},\ }\href@noop {} {\bibfield  {journal} {\bibinfo  {journal} {J.
  Stat. Phys.}\ }\textbf {\bibinfo {volume} {100}},\ \bibinfo {pages} {523}
  (\bibinfo {year} {2000})}\BibitemShut {NoStop}%
\bibitem [{\citenamefont {Pr{\"a}hofer}\ and\ \citenamefont
  {Spohn}(2002)}]{PS2002a}%
  \BibitemOpen
  \bibfield  {author} {\bibinfo {author} {\bibfnamefont {M.}~\bibnamefont
  {Pr{\"a}hofer}}\ and\ \bibinfo {author} {\bibfnamefont {H.}~\bibnamefont
  {Spohn}},\ }in\ \href@noop {} {\emph {\bibinfo {booktitle} {In and out of
  equilibrium, vol. 51 of {\it Progress in Probability}}}},\ \bibinfo {editor}
  {edited by\ \bibinfo {editor} {\bibfnamefont {V.}~\bibnamefont
  {Sidoravicius}}}\ (\bibinfo {year} {2002})\ pp.\ \bibinfo {pages}
  {185--204}\BibitemShut {NoStop}%
\bibitem [{\citenamefont {Sasamoto}(2005)}]{Sasamoto2005}%
  \BibitemOpen
  \bibfield  {author} {\bibinfo {author} {\bibfnamefont {T.}~\bibnamefont
  {Sasamoto}},\ }\href@noop {} {\bibfield  {journal} {\bibinfo  {journal} {J.
  Phys. A}\ }\textbf {\bibinfo {volume} {38}},\ \bibinfo {pages} {L549}
  (\bibinfo {year} {2005})}\BibitemShut {NoStop}%
\bibitem [{\citenamefont {Calabrese}\ and\ \citenamefont
  {Doussal}(2011)}]{CLD2011}%
  \BibitemOpen
  \bibfield  {author} {\bibinfo {author} {\bibfnamefont {P.}~\bibnamefont
  {Calabrese}}\ and\ \bibinfo {author} {\bibfnamefont {P.~L.}\ \bibnamefont
  {Doussal}},\ }\href@noop {} {\bibfield  {journal} {\bibinfo  {journal} {Phys.
  Rev. Lett.}\ }\textbf {\bibinfo {volume} {106}},\ \bibinfo {pages} {250603}
  (\bibinfo {year} {2011})}\BibitemShut {NoStop}%
\bibitem [{\citenamefont {{K. A. Takeuchi and M. Sano}}(2010)}]{TS2010}%
  \BibitemOpen
  \bibfield  {author} {\bibinfo {author} {\bibnamefont {{K. A. Takeuchi and M.
  Sano}}},\ }\href@noop {} {\bibfield  {journal} {\bibinfo  {journal} {Phys.
  Rev. Lett.}\ }\textbf {\bibinfo {volume} {104}},\ \bibinfo {pages} {230601}
  (\bibinfo {year} {2010})}\BibitemShut {NoStop}%
\bibitem [{\citenamefont {{K. A. Takeuchi and M. Sano}}(2012)}]{TS2012}%
  \BibitemOpen
  \bibfield  {author} {\bibinfo {author} {\bibnamefont {{K. A. Takeuchi and M.
  Sano}}},\ }\href@noop {} {\bibfield  {journal} {\bibinfo  {journal} {J. Stat.
  Phys.}\ }\textbf {\bibinfo {volume} {147}},\ \bibinfo {pages} {853} (\bibinfo
  {year} {2012})}\BibitemShut {NoStop}%
\bibitem [{\citenamefont {{K. A. Takeuch, M. Sano, T. Sasamoto and H.
  Spohn}}(2011)}]{TSSS2011}%
  \BibitemOpen
  \bibfield  {author} {\bibinfo {author} {\bibnamefont {{K. A. Takeuch, M.
  Sano, T. Sasamoto and H. Spohn}}},\ }\href@noop {} {\bibfield  {journal}
  {\bibinfo  {journal} {Sci. Pep.}\ }\textbf {\bibinfo {volume} {1}},\ \bibinfo
  {pages} {34} (\bibinfo {year} {2011})}\BibitemShut {NoStop}%
\bibitem [{\citenamefont {{Y. T. Fukai, K. A. Takeuchi}}(2017)}]{FT2017}%
  \BibitemOpen
  \bibfield  {author} {\bibinfo {author} {\bibnamefont {{Y. T. Fukai, K. A.
  Takeuchi}}},\ }\href@noop {} {\bibfield  {journal} {\bibinfo  {journal}
  {Phys. Rev. Lett.}\ }\textbf {\bibinfo {volume} {119}},\ \bibinfo {pages}
  {030602} (\bibinfo {year} {2017})}\BibitemShut {NoStop}%
\bibitem [{\citenamefont {{R. A. Blythe, M. R. Evans}}(2007)}]{BE2007}%
  \BibitemOpen
  \bibfield  {author} {\bibinfo {author} {\bibnamefont {{R. A. Blythe, M. R.
  Evans}}},\ }\href@noop {} {\bibfield  {journal} {\bibinfo  {journal} {J.
  Phys. A: Math. Theor.}\ }\textbf {\bibinfo {volume} {40}},\ \bibinfo {pages}
  {R333} (\bibinfo {year} {2007})}\BibitemShut {NoStop}%
\bibitem [{\citenamefont {{S. Prolhac, M. R. Evans, K.
  Mallick}}(2009)}]{PEM2009}%
  \BibitemOpen
  \bibfield  {author} {\bibinfo {author} {\bibnamefont {{S. Prolhac, M. R.
  Evans, K. Mallick}}},\ }\href@noop {} {\bibfield  {journal} {\bibinfo
  {journal} {J. Phys. A}\ }\textbf {\bibinfo {volume} {42}},\ \bibinfo {pages}
  {165004} (\bibinfo {year} {2009})}\BibitemShut {NoStop}%
\bibitem [{\citenamefont {{A. Kuniba, S. Maruyama and M.
  Okado}}(2016)}]{KMO2016}%
  \BibitemOpen
  \bibfield  {author} {\bibinfo {author} {\bibnamefont {{A. Kuniba, S. Maruyama
  and M. Okado}}},\ }\href@noop {} {\bibfield  {journal} {\bibinfo  {journal}
  {J. Int. Systems}\ }\textbf {\bibinfo {volume} {1}},\ \bibinfo {pages}
  {xyw002} (\bibinfo {year} {2016})}\BibitemShut {NoStop}%
\bibitem [{\citenamefont {{N. Crampe, M. R. Evans, K. Mallick, E. Ragoucy, M.
  Vanicat}}(2016)}]{CEMRV2016}%
  \BibitemOpen
  \bibfield  {author} {\bibinfo {author} {\bibnamefont {{N. Crampe, M. R.
  Evans, K. Mallick, E. Ragoucy, M. Vanicat}}},\ }\href@noop {} {\bibfield
  {journal} {\bibinfo  {journal} {J. Phys. A: Math. Theor.}\ }\textbf {\bibinfo
  {volume} {49}},\ \bibinfo {pages} {475001} (\bibinfo {year}
  {2016})}\BibitemShut {NoStop}%
\bibitem [{\citenamefont {{F. C. Alcaraz, M. Droz, M. Henkel, and V.
  Rittenberg}}(1994)}]{ADHR1994}%
  \BibitemOpen
  \bibfield  {author} {\bibinfo {author} {\bibnamefont {{F. C. Alcaraz, M.
  Droz, M. Henkel, and V. Rittenberg}}},\ }\href@noop {} {\bibfield  {journal}
  {\bibinfo  {journal} {Ann. Phys.}\ }\textbf {\bibinfo {volume} {230}},\
  \bibinfo {pages} {250} (\bibinfo {year} {1994})}\BibitemShut {NoStop}%
\bibitem [{\citenamefont {Kim}\ and\ \citenamefont {den
  Nijs}(2007)}]{KimdenNijs2007}%
  \BibitemOpen
  \bibfield  {author} {\bibinfo {author} {\bibfnamefont {K.~H.}\ \bibnamefont
  {Kim}}\ and\ \bibinfo {author} {\bibfnamefont {M.}~\bibnamefont {den Nijs}},\
  }\href@noop {} {\bibfield  {journal} {\bibinfo  {journal} {Phys. Rev. E.}\
  }\textbf {\bibinfo {volume} {76}},\ \bibinfo {pages} {21107} (\bibinfo {year}
  {2007})}\BibitemShut {NoStop}%
\bibitem [{\citenamefont {{C. Arita, A. Kuniba, K. Sakai, T.
  Sawabe}}(2009)}]{AKSS2009}%
  \BibitemOpen
  \bibfield  {author} {\bibinfo {author} {\bibnamefont {{C. Arita, A. Kuniba,
  K. Sakai, T. Sawabe}}},\ }\href@noop {} {\bibfield  {journal} {\bibinfo
  {journal} {J. Phys. A: Math. Theor.}\ }\textbf {\bibinfo {volume} {42}},\
  \bibinfo {pages} {345002} (\bibinfo {year} {2009})}\BibitemShut {NoStop}%
\bibitem [{\citenamefont {{C. A. Tracy and H. Widom}}(2013)}]{TW2013}%
  \BibitemOpen
  \bibfield  {author} {\bibinfo {author} {\bibnamefont {{C. A. Tracy and H.
  Widom}}},\ }\href@noop {} {\bibfield  {journal} {\bibinfo  {journal} {J.
  Stat. Phys.}\ }\textbf {\bibinfo {volume} {150}},\ \bibinfo {pages} {457}
  (\bibinfo {year} {2013})}\BibitemShut {NoStop}%
\bibitem [{\citenamefont {{J. Kuan}}()}]{Kuan2018p}%
  \BibitemOpen
  \bibfield  {author} {\bibinfo {author} {\bibnamefont {{J. Kuan}}},\
  }\href@noop {} {\ }\Eprint {http://arxiv.org/abs/arXiv:1801.02313}
  {arXiv:1801.02313} \BibitemShut {NoStop}%
\bibitem [{\citenamefont {{J. Kuan}}(2016)}]{Kuan2016}%
  \BibitemOpen
  \bibfield  {author} {\bibinfo {author} {\bibnamefont {{J. Kuan}}},\
  }\href@noop {} {\bibfield  {journal} {\bibinfo  {journal} {J. Phys. A: Math.
  Theor.}\ }\textbf {\bibinfo {volume} {49}},\ \bibinfo {pages} {115002}
  (\bibinfo {year} {2016})}\BibitemShut {NoStop}%
\bibitem [{\citenamefont {{Z. Chen, J. de Gier, M. Wheeler}}()}]{CdGW2017p}%
  \BibitemOpen
  \bibfield  {author} {\bibinfo {author} {\bibnamefont {{Z. Chen, J. de Gier,
  M. Wheeler}}},\ }\href@noop {} {\ }\Eprint
  {http://arxiv.org/abs/arXiv:1709.06227} {arXiv:1709.06227} \BibitemShut
  {NoStop}%
\bibitem [{\citenamefont {{P. L. Ferrari, P. Nejjar, P. Ghosal}}()}]{FNG2017p}%
  \BibitemOpen
  \bibfield  {author} {\bibinfo {author} {\bibnamefont {{P. L. Ferrari, P.
  Nejjar, P. Ghosal}}},\ }\href@noop {} {\ }\Eprint
  {http://arxiv.org/abs/arXiv:1710.02323} {arXiv:1710.02323} \BibitemShut
  {NoStop}%
\bibitem [{\citenamefont {{P. F. Arndt, T. Heinzel, V.
  Rittenberg}}(1999)}]{AHR1999}%
  \BibitemOpen
  \bibfield  {author} {\bibinfo {author} {\bibnamefont {{P. F. Arndt, T.
  Heinzel, V. Rittenberg}}},\ }\href@noop {} {\bibfield  {journal} {\bibinfo
  {journal} {J. Stat. Phys.}\ }\textbf {\bibinfo {volume} {97}},\ \bibinfo
  {pages} {1} (\bibinfo {year} {1999})}\BibitemShut {NoStop}%
\bibitem [{\citenamefont {Tracy}\ and\ \citenamefont {Widom}(1994)}]{TW1994}%
  \BibitemOpen
  \bibfield  {author} {\bibinfo {author} {\bibfnamefont {C.~A.}\ \bibnamefont
  {Tracy}}\ and\ \bibinfo {author} {\bibfnamefont {H.}~\bibnamefont {Widom}},\
  }\href@noop {} {\bibfield  {journal} {\bibinfo  {journal} {Comm. Math.
  Phys.}\ }\textbf {\bibinfo {volume} {159}},\ \bibinfo {pages} {151} (\bibinfo
  {year} {1994})}\BibitemShut {NoStop}%
\bibitem [{\citenamefont {Mehta}(2004)}]{Mehta2004}%
  \BibitemOpen
  \bibfield  {author} {\bibinfo {author} {\bibfnamefont {M.~L.}\ \bibnamefont
  {Mehta}},\ }\href@noop {} {\emph {\bibinfo {title} {Random Matrices}}},\
  \bibinfo {edition} {3rd}\ ed.\ (\bibinfo  {publisher} {Elsevier},\ \bibinfo
  {year} {2004})\BibitemShut {NoStop}%
\bibitem [{\citenamefont {Forrester}(2010)}]{Forrester2010}%
  \BibitemOpen
  \bibfield  {author} {\bibinfo {author} {\bibfnamefont {P.~J.}\ \bibnamefont
  {Forrester}},\ }\href@noop {} {\emph {\bibinfo {title} {Log gases and random
  matrices}}}\ (\bibinfo  {publisher} {Princeton University Press},\ \bibinfo
  {year} {2010})\BibitemShut {NoStop}%
\bibitem [{\citenamefont {{L. Cantini}}(2008)}]{Cantini2008}%
  \BibitemOpen
  \bibfield  {author} {\bibinfo {author} {\bibnamefont {{L. Cantini}}},\
  }\href@noop {} {\bibfield  {journal} {\bibinfo  {journal} {J. Phys. A.}\
  }\textbf {\bibinfo {volume} {41}},\ \bibinfo {pages} {095001} (\bibinfo
  {year} {2008})}\BibitemShut {NoStop}%
\bibitem [{\citenamefont {{N. Rajewsky, T. Sasamoto and E. R.
  Speer}}(2000)}]{RSS2000}%
  \BibitemOpen
  \bibfield  {author} {\bibinfo {author} {\bibnamefont {{N. Rajewsky, T.
  Sasamoto and E. R. Speer}}},\ }\href@noop {} {\bibfield  {journal} {\bibinfo
  {journal} {Physica. A}\ }\textbf {\bibinfo {volume} {279}},\ \bibinfo {pages}
  {123} (\bibinfo {year} {2000})}\BibitemShut {NoStop}%
\bibitem [{\citenamefont {{Z. Chen, J. de Gier, I. Hiki, T.
  Sasamoto}}(2018)}]{ChenGHS}%
  \BibitemOpen
  \bibfield  {author} {\bibinfo {author} {\bibnamefont {{Z. Chen, J. de Gier,
  I. Hiki, T. Sasamoto}}},\ }\href@noop {} {\bibfield  {journal} {\bibinfo
  {journal} {{in preparation}}\ } (\bibinfo {year} {2018})}\BibitemShut
  {NoStop}%
\bibitem [{\citenamefont {Sch{\"u}tz}(1997)}]{Schuetz1997b}%
  \BibitemOpen
  \bibfield  {author} {\bibinfo {author} {\bibfnamefont {G.~M.}\ \bibnamefont
  {Sch{\"u}tz}},\ }\href@noop {} {\bibfield  {journal} {\bibinfo  {journal} {J.
  Stat. Phys.}\ }\textbf {\bibinfo {volume} {88}},\ \bibinfo {pages} {427}
  (\bibinfo {year} {1997})}\BibitemShut {NoStop}%
\bibitem [{\citenamefont {Widom}(1993)}]{widom}%
  \BibitemOpen
  \bibfield  {author} {\bibinfo {author} {\bibfnamefont {M.}~\bibnamefont
  {Widom}},\ }\href {\doibase 10.1103/PhysRevLett.70.2094} {\bibfield
  {journal} {\bibinfo  {journal} {Phys. Rev. Lett.}\ }\textbf {\bibinfo
  {volume} {70}},\ \bibinfo {pages} {2094} (\bibinfo {year}
  {1993})}\BibitemShut {NoStop}%
\bibitem [{\citenamefont {de~Gier}\ and\ \citenamefont
  {Nienhuis}(1997)}]{gierN97}%
  \BibitemOpen
  \bibfield  {author} {\bibinfo {author} {\bibfnamefont {J.}~\bibnamefont
  {de~Gier}}\ and\ \bibinfo {author} {\bibfnamefont {B.}~\bibnamefont
  {Nienhuis}},\ }\href {\doibase 10.1007/BF02181494} {\bibfield  {journal}
  {\bibinfo  {journal} {J. Stat. Phys.}\ }\textbf {\bibinfo {volume} {87}},\
  \bibinfo {pages} {415} (\bibinfo {year} {1997})}\BibitemShut {NoStop}%
\bibitem [{\citenamefont {de~Gier}\ and\ \citenamefont
  {Nienhuis}(1998)}]{gierN98}%
  \BibitemOpen
  \bibfield  {author} {\bibinfo {author} {\bibfnamefont {J.}~\bibnamefont
  {de~Gier}}\ and\ \bibinfo {author} {\bibfnamefont {B.}~\bibnamefont
  {Nienhuis}},\ }\href {http://stacks.iop.org/0305-4470/31/i=9/a=006}
  {\bibfield  {journal} {\bibinfo  {journal} {J. Phys. A}\ }\textbf {\bibinfo
  {volume} {31}},\ \bibinfo {pages} {2141} (\bibinfo {year}
  {1998})}\BibitemShut {NoStop}%
\bibitem [{\citenamefont {{P. Zinn-Justin}}(2009)}]{ZinnJustin2009}%
  \BibitemOpen
  \bibfield  {author} {\bibinfo {author} {\bibnamefont {{P. Zinn-Justin}}},\
  }\href@noop {} {\bibfield  {journal} {\bibinfo  {journal} {E. J. Combin.}\
  }\textbf {\bibinfo {volume} {16}},\ \bibinfo {pages} {1745} (\bibinfo {year}
  {2009})},\ \Eprint {http://arxiv.org/abs/arXiv:0809.2392} {arXiv:0809.2392}
  \BibitemShut {NoStop}%
\bibitem [{\citenamefont {{M. Wheeler and P. Zinn-Justin}}()}]{WZ2017}%
  \BibitemOpen
  \bibfield  {author} {\bibinfo {author} {\bibnamefont {{M. Wheeler and P.
  Zinn-Justin}}},\ }\href@noop {} {\bibfield  {journal} {\bibinfo  {journal}
  {J. Reine Angew. Math. (Crelle)}\ }}\Eprint {http://arxiv.org/abs/published
  online 2017-09-21} {published online 2017-09-21} \BibitemShut {NoStop}%
\bibitem [{\citenamefont {{S. Kharchev, A. Marshakov, A. Mironov, A. Morozov,
  S. Pakuliak}}(1993)}]{KMMMP1993}%
  \BibitemOpen
  \bibfield  {author} {\bibinfo {author} {\bibnamefont {{S. Kharchev, A.
  Marshakov, A. Mironov, A. Morozov, S. Pakuliak}}},\ }\href@noop {} {\bibfield
   {journal} {\bibinfo  {journal} {Nucl. Phys. B.}\ }\textbf {\bibinfo {volume}
  {404}},\ \bibinfo {pages} {717} (\bibinfo {year} {1993})}\BibitemShut
  {NoStop}%
\bibitem [{\citenamefont {Kostov}(1996)}]{Kostov1996}%
  \BibitemOpen
  \bibfield  {author} {\bibinfo {author} {\bibfnamefont {I.}~\bibnamefont
  {Kostov}},\ }\href@noop {} {\bibfield  {journal} {\bibinfo  {journal} {Nucl.
  Rev. B.}\ }\textbf {\bibinfo {volume} {45A}},\ \bibinfo {pages} {13}
  (\bibinfo {year} {1996})}\BibitemShut {NoStop}%
\bibitem [{\citenamefont {{J. Fritz, B. T\'oth}}(2004)}]{FT2004}%
  \BibitemOpen
  \bibfield  {author} {\bibinfo {author} {\bibnamefont {{J. Fritz, B.
  T\'oth}}},\ }\href@noop {} {\bibfield  {journal} {\bibinfo  {journal} {Comm.
  Math. Phys.}\ }\textbf {\bibinfo {volume} {249}},\ \bibinfo {pages} {1}
  (\bibinfo {year} {2004})}\BibitemShut {NoStop}%
\bibitem [{\citenamefont {Bressan}(2013)}]{Bressan2009}%
  \BibitemOpen
  \bibfield  {author} {\bibinfo {author} {\bibfnamefont {A.}~\bibnamefont
  {Bressan}},\ }in\ \href@noop {} {\emph {\bibinfo {booktitle} {Modelling and
  Optimisation of Flows on Networks, Cetraro, Italy 2009. {\it Lecture Notes in
  Mathematics 2062}}}},\ \bibinfo {editor} {edited by\ \bibinfo {editor}
  {\bibfnamefont {M.~R.~e.}\ \bibnamefont {P.~Benedetto}}}\ (\bibinfo {year}
  {2013})\ pp.\ \bibinfo {pages} {157--245}\BibitemShut {NoStop}%
\bibitem [{\citenamefont {{D. Erta\c{s}, M. Kardar}}(1992)}]{EK1992}%
  \BibitemOpen
  \bibfield  {author} {\bibinfo {author} {\bibnamefont {{D. Erta\c{s}, M.
  Kardar}}},\ }\href@noop {} {\bibfield  {journal} {\bibinfo  {journal} {Phys.
  Rev. Lett.}\ }\textbf {\bibinfo {volume} {69}},\ \bibinfo {pages} {929}
  (\bibinfo {year} {1992})}\BibitemShut {NoStop}%
\bibitem [{\citenamefont {{T. Funaki and M. Hoshino}}(2016)}]{FH2016}%
  \BibitemOpen
  \bibfield  {author} {\bibinfo {author} {\bibnamefont {{T. Funaki and M.
  Hoshino}}},\ }\href@noop {} {\bibfield  {journal} {\bibinfo  {journal} {J.
  Func. Anal.}\ }\textbf {\bibinfo {volume} {273}},\ \bibinfo {pages} {1165}
  (\bibinfo {year} {2016})}\BibitemShut {NoStop}%
\bibitem [{\citenamefont {Imamura}\ and\ \citenamefont {Sasamoto}()}]{IS2017p}%
  \BibitemOpen
  \bibfield  {author} {\bibinfo {author} {\bibfnamefont {T.}~\bibnamefont
  {Imamura}}\ and\ \bibinfo {author} {\bibfnamefont {T.}~\bibnamefont
  {Sasamoto}},\ }\href@noop {} {\ }\Eprint
  {http://arxiv.org/abs/arXiv:1701.05991} {arXiv:1701.05991} \BibitemShut
  {NoStop}%
\bibitem [{\citenamefont {Borodin}\ \emph {et~al.}(2007)\citenamefont
  {Borodin}, \citenamefont {Ferrari},\ and\ \citenamefont
  {Pr{\"a}hofer}}]{BFP2007}%
  \BibitemOpen
  \bibfield  {author} {\bibinfo {author} {\bibfnamefont {A.}~\bibnamefont
  {Borodin}}, \bibinfo {author} {\bibfnamefont {P.~L.}\ \bibnamefont
  {Ferrari}}, \ and\ \bibinfo {author} {\bibfnamefont {M.}~\bibnamefont
  {Pr{\"a}hofer}},\ }\href@noop {} {\bibfield  {journal} {\bibinfo  {journal}
  {Int. Math. Res. Papers}\ }\textbf {\bibinfo {volume} {2007}},\ \bibinfo
  {pages} {rpm002} (\bibinfo {year} {2007})}\BibitemShut {NoStop}%
\end{thebibliography}
\end{document}